\documentclass[reprint,superscriptaddress,preprintnumbers,amsmath,amssymb,aps,prd,floatfix]{revtex4-2}
\usepackage{graphicx}

\usepackage[skip=-5pt,font=small,justification=RaggedRight]{caption}

\def\babar{B{\sc a}B{\sc ar}~}
\def\beq{\begin{equation}}
\def\eeq{\end{equation}}
\def\bea{\begin{eqnarray}}
\def\eea{\end{eqnarray}}
\def\nn{\nonumber}
\def\bd{B^0}
\def\bs{B_s^0}
\def\btod{{\bar b} \to {\bar d}}
\def\btos{{\bar b} \to {\bar s}}
\def\btokpipi{B \to K \pi \pi}
\def\btokkk{B \to KK{\bar K}}
\def \ok{\overline{K}^0}
\def \FA{{(fa)}}
\def \ll{\left|}
\def \rr{\right|}
\def \<{\left<}
\def \>{\right>}
\def \[{\left[}
\def \]{\right]}
\def \({\left(}
\def \){\right)}
\def \lb{\left\{}
\def \rb{\right\}}
\def \nl{\nonumber\\}
\def \s{\sqrt{2}}
\def \st{\sqrt{3}}
\def \sf{\sqrt{5}}
\def \sx{\sqrt{6}}
\def \ssot{\frac{\s}{\st}}
\def \ssof{\frac{\s}{\sf}}
\def \ssofi{\frac{\s}{\sqrt{15}}}
\def\oof{\frac{1}{\sf}}
\def\oofi{\frac{1}{\sqrt{15}}}
\def\oot{\frac{1}{\st}}
\def \hf{\frac{1}{2}}
\def \thf{\frac{3}{2}}
\def \oth{\frac{1}{3}}
\def \tth{\frac{2}{3}}
\def \ons{\frac{1}{\s}}
\def \ths{\frac{3}{\s}}
\def\text#1{{\rm #1}}
\def\degree{{^\circ}}
\def\cA{{\cal A}}

\allowdisplaybreaks
\begin{document}

\preprint{UdeM-GPP-TH-23-298}

\title{\boldmath Charmless $B\to PPP$ Decays: \\ the Fully-Antisymmetric Final State}

\author{Bhubanjyoti Bhattacharya}
\email{bbhattach@ltu.edu}
\affiliation{Department of Natural Sciences, Lawrence Technological University, Southfield, MI 48075, USA}

\author{Mirjam Fines-Neuschild}
\email{mirjam.fines-neuschild@mail.concordia.ca}
\affiliation{Centre for Engineering in Society (CES), Concordia University, 1515 Ste-Catherine St.\ W., Montr\'eal, QC, Canada H3G 2W1}

\author{Andrea~Houck}
\email{ahouck@ltu.edu}
\affiliation{Department of Natural Sciences, Lawrence Technological University, Southfield, MI 48075, USA}

\author{Maxime Imbeault}
\email{mimbeault@cegepsl.qc.ca}
\affiliation{D\'epartement de physique, C\'egep de Saint-Laurent, 625, avenue Sainte-Croix, Montr\'eal, QC, Canada H4L 3X7}

\author{Alexandre Jean}
\email{alexandre.jean.1@umontreal.ca}
\affiliation{Physique des Particules, Universit\'e de Montr\'eal, 1375 Avenue Th\'er\`ese-Lavoie-Roux, Montr\'eal, QC, Canada  H2V 0B3}

\author{David London}
\email{london@lps.umontreal.ca}
\affiliation{Physique des Particules, Universit\'e de Montr\'eal, 1375 Avenue Th\'er\`ese-Lavoie-Roux, Montr\'eal, QC, Canada  H2V 0B3}

\begin{abstract}
Under flavor $SU(3)$ symmetry (SU(3)$_F$), the final-state particles in $B\to PPP$ decays ($P$ is a pseudoscalar meson) are treated as identical, and the $PPP$ must be in a fully-symmetric (FS) state, a fully-antisymmetric (FA) state, or in one of four mixed states. In this paper, we present the formalism for the FA states. We write the amplitudes for the 22 $B\to PPP$ decays that can be in an FA state in terms of both SU(3)$_F$ reduced matrix elements and diagrams. This shows the equivalence of diagrams and SU(3)$_F$. We also give 15 relations among the amplitudes in the SU(3)$_F$ limit, as well as the additional four that appear when the diagrams $E$/$A$/$PA$ are neglected. We present sets of $B \to PPP$ decays that can be used to extract $\gamma$ using the FA amplitudes. The value(s) of $\gamma$ found in this way can be compared with the value(s) found using the FS states.
\end{abstract}

\maketitle

\section{Introduction}

The $B$-factories \babar and Belle were built with the goal of measuring CP violation in $B$ decays. The idea was to measure the three angles of the unitarity triangle, $\alpha$, $\beta$, and $\gamma$, and to test the standard model (SM) by seeing if $\alpha + \beta + \gamma = \pi$. Now, $\alpha$ is measured using $B \to \pi\pi$ decays, and the (loop-level) penguin contribution is removed using an isospin analysis \cite{Gronau:1990ka}. $\beta$ is mainly measured in decays such as $B^0 \to J/\psi K_S$, which are dominated by the tree contribution. And the standard methods of measuring $\gamma$ \cite{Gronau:1990ra, Gronau:1991dp, Atwood:1996ci, Giri:2003ty} involve only tree-level decays. As a result, NP can affect these measurements only if it can compete with the tree-level SM contributions. (In principle, there could be (loop-level) NP contributions to $B^0$-${\bar B}^0$ mixing, but these effects cancel in the sum $\alpha + \beta$ \cite{Nir:1990yq}.) Given that no new particles have been seen at the LHC, we now know that the NP must be heavy, so its contributions cannot compete with those of the SM at tree level. It is therefore unsurprising that $\alpha + \beta + \gamma \simeq \pi$ was found \cite{HeavyFlavorAveragingGroup:2022wzx}.

Another way to search for NP using CP violation in $B$ decays is to measure the same CP phase in two different ways. If the results do not agree, this would reveal the presence of NP. An example is $\beta$. At the quark level, the decay $B^0 \to J/\psi K_S$ is ${\bar b} \to {\bar c} c {\bar s}$, which has no weak phase in the SM. Similarly, the decay $B^0 \to \phi K_S$ involves ${\bar b} \to {\bar s} s {\bar s}$, which can arise only via loop-level gluonic and electroweak penguin contributions, and also has no weak phase in the SM, to a good approximation. The point is that $\beta$ can be measured using either decay \cite{London:1997zk}. The difference between the two is that, while tree-level NP contributions are much smaller than tree-level SM contributions, they {\it can} be of the same order as loop-level SM contributions. Thus, a difference between the (tree-level) measurement of $\beta$ in $B^0 \to J/\psi K_S$ and its (loop-level) measurement in $B^0 \to \phi K_S$ would point to a (tree-level) NP contribution to ${\bar b} \to {\bar s} s {\bar s}$. Experiments have searched for such a discrepancy, but none has been observed \cite{HeavyFlavorAveragingGroup:2022wzx}.

In principle, this can also be done with $\gamma$. If $\gamma$ could be extracted from decays that receive significant penguin contributions (gluonic and/or electroweak), one could compare this (loop-level) measurement of $\gamma$ with that of the (tree-level) methods of Refs.~\cite{Gronau:1990ra, Gronau:1991dp, Atwood:1996ci, Giri:2003ty}.

In fact, methods for making a loop-level measurement of $\gamma$ were proposed in Refs.~\cite{Imbeault:2010xg, Rey-LeLorier:2011ltd, Bhattacharya:2015uua}. They all involve charmless, three-body $B\to PPP$ decays ($P$ is a pseudoscalar meson). Under flavor SU(3) symmetry (SU(3)$_F$), the three final-state
particles are treated as identical. The total final-state wavefunction must be symmetric,
so that the six permutations of these particles must be considered:
the $PPP$ must be in a fully-symmetric state, a
fully-antisymmetric state, or in one of four mixed states under SU(3)$_F$.

For the measurement of the decay $B \to P_1 P_2 P_3$, the results are
usually presented in the form of a Dalitz plot. This is a function of
two of the three Mandelstam variables, say $s_{12}$ and $s_{13}$, where $s_{ij} \equiv \left( p_i + p_j
\right)^2$. One can then perform an isobar
analysis, which is essentially a fit of the Dalitz plot to a
non-resonant and various intermediate resonant contributions, to obtain the
decay amplitude ${\cal M} (s_{12}, s_{13})$ describing $B \to P_1 P_2
P_3$. In Ref.~\cite{Lorier:2010xf}, it is pointed out that one can
use ${\cal M} (s_{12}, s_{13})$ to construct the amplitudes for the
individual fully-symmetric,
fully-antisymmetric and mixed final states. In this way, one can study
decays into final states with each of the possible symmetries.

Ref.~\cite{Lorier:2010xf} also shows that the $B \to PPP$ amplitudes can be written in terms of diagrams similar to those used in $B \to PP$ decays \cite{Gronau:1994rj, Gronau:1995hn}. The main advantage of using diagrams to describe $B$-decay amplitudes is that it can be argued on dynamical grounds that certain diagrams are subdominant. The neglect of these diagrams greatly simplifies the analysis. We note that, for $B \to PP$ decays, this theoretical assumption has been borne out by experiment: decays that are mediated by these supposedly subdominant diagrams, such as $B^0 \to K^+ K^-$ and $B_s^0 \to \pi^+ \pi^-$, are indeed found to have branching ratios considerably smaller than those of other charmless $B \to PP$ decays.

Still, we stress that this assumption does not follow from group theory. Before putting it into practice, it must be shown that the description of the amplitudes using the full set of diagrams is equivalent to a description in terms of SU(3)$_F$ reduced matrix elements (RMEs). In Ref.~\cite{Bhattacharya:2014eca}, this is demonstrated explicitly for the fully-symmetric (FS) final state in $B \to PPP$. It is therefore justified to use a diagrammatic description of these decay amplitudes and to neglect certain diagrams.

In the methods proposed in Refs.~\cite{Imbeault:2010xg, Rey-LeLorier:2011ltd, Bhattacharya:2015uua}, these techniques are used to cleanly extract the weak phase $\gamma$ from the FS states of various $B \to PPP$ decays. The method of Ref.~\cite{Rey-LeLorier:2011ltd} is particularly interesting. It
combines information from the Dalitz plots for $\bd \to
K^+\pi^0\pi^-$, $\bd \to K^0\pi^+\pi^-$, $B^+ \to K^+\pi^+\pi^-$, $\bd
\to K^+ K^0 K^-$, and $\bd \to K^0 K^0 \ok$. These
$\btokpipi$ and $\btokkk$ decays all receive loop-level penguin and
electroweak-penguin contributions, so it is a loop-level value of $\gamma$ that is measured here. As noted above, the comparison of the tree-level and
loop-level measurements of $\gamma$ is an excellent test of the
Standard Model.

This method was applied in Ref.~\cite{Bhattacharya:2013cla} to the
measurements of the Dalitz plots of the five $\btokpipi$ and $\btokkk$
decays by the \babar\ Collaboration \cite{BaBar:2011vfx,
  BaBar:2009jov, BaBar:2008lpx, BaBar:2012iuj, BaBar:2011ktx}.
However, this was a theoretical analysis: by its own
admission, it did not properly take all the errors into account. This
was improved in Ref.~\cite{Bertholet:2018tmx}, which was a
collaboration of theory and experiment. Six possible values of
$\gamma$ were found:
\bea
\gamma_1 & = & [ \phantom{1} 12.9\mbox{}\,^{+8.4\phantom{1}}_{-4.3}   \text{ (stat)}  \pm 1.3 \text{ (syst)}]\degree ~,\nl
\gamma_2 & = & [ \phantom{1} 36.6\mbox{}\,^{+6.6\phantom{1}}_{-6.1}  \text{ (stat)}  \pm  2.6 \text{ (syst)}]\degree ~,\nl
\gamma_3 & = & [ \phantom{1} 68.9\mbox{}\,^{+8.6\phantom{1}}_{-8.6}  \text{ (stat)}  \pm  2.4 \text{ (syst)}]\degree ~,\nl
\gamma_4 & = & [ 223.2\mbox{}\,^{+10.9}_{-7.5}  \text{ (stat)} \pm  1.0 \text{ (syst)}]\degree ~,\nl
\gamma_5 & = & [ 266.4\mbox{}\,^{+9.2}_{-10.8}  \text{ (stat)} \pm  1.9 \text{ (syst)}]\degree ~,\nl
\gamma_6 & = & [ 307.5\mbox{}\,^{+6.9\phantom{1}}_{-8.1}  \text{ (stat)}  \pm  1.1 \text{ (syst)}]\degree ~.
\eea
One solution -- $\gamma_3$ -- is compatible with the latest world
average tree-level value, $\gamma = (66.2^{+3.4}_{-3.6})^\circ$
\cite{HeavyFlavorAveragingGroup:2022wzx}. The other solutions are in disagreement, perhaps
hinting at new physics. In addition, it is found that, when averaged
over the entire Dalitz plane, the effect of SU(3)$_F$ breaking on the
analysis is only at the percent level.

At this stage, the burning question is: what is the true value of
$\gamma$ in this system? The above analysis was carried out using the
FS final state. One way this question might be answered is to repeat
the analysis -- or perform a different analysis to extract $\gamma$ -- using a different symmetry of the final state. The hope is
that, if there are again multiple solutions for $\gamma$, only one
will be common to the two sets of solutions; this will be the true
value of $\gamma$. And if it differs from the tree-level value, this
will be a smoking-gun signal of new physics.

The formalism describing $B \to PPP$ decays with a FS final state was presented in Refs.~\cite{Lorier:2010xf,Imbeault:2010xg,Rey-LeLorier:2011ltd, Bhattacharya:2014eca}. However, the same formalism has not been given for the other final-state symmetries. In this paper, we focus on $B \to PPP$ decays in which the final state is fully antisymmetric.

We begin in Sec.~2 with a presentation of the Wigner-Eckart decomposition of the FA $B \to PPP$ amplitudes in terms of SU(3)$_F$ reduced matrix elements. A similar decomposition in terms of diagrams is given in Sec,~3, thereby demonstrating the equivalence of SU(3)$_F$ reduced matrix elements and diagrams. Various relations among the amplitude are given in Sec.~4. Sec.~5 discusses the consequences of neglecting the $E$/$A$/$PA$ diagrams, which are expected to be smaller than the other diagrams. Various applications of this formalism, including the extraction of $\gamma$ and the measurement of SU(3)$_F$ breaking,  are elaborated in Sec.~6. We conclude in Sec.~7.

\section{\boldmath SU(3)$_F$ Wigner-Eckart Decomposition} \label{sec:RMEs}

We begin by representing the $B\to PPP$ decay amplitudes for fully-antisymmetric (FA) final states in terms of SU(3)$_F$ reduced matrix elements. The amplitude for a decay process involves three pieces: a) the initial state, b) the Hamiltonian, and c) the final state. Here, the SU(3)$_F$ representations of the decaying $B$ mesons and the underlying quark-level transitions are identical to those used in Ref.~\cite{Bhattacharya:2014eca}, where the FS state was studied. The three-body final states we consider in this article are new: under the exchange of any two of the three final-state particles, the $\ll PPP\>$ states considered in this article are fully antisymmetric.

In this section, we perform SU(3)$_F$ Wigner-Eckart decompositions of the FA $B \to PPP$ decay amplitudes. We adopt the notation used in Ref.~\cite{Bhattacharya:2014eca} and represent each element of SU(3)$_F$ by $\ll{\bf r} Y I I_3\>$, where ${\bf r}$ is the irreducible
representation (irrep) of SU(3)$_F$, $Y$ is the hypercharge, and $I$ and $I_3$
stand for the isospin and its third component, respectively. Note that, in
general, Lie algebras are not associative, so that the order of
multiplication of elements is important. Here we take products from left to right. We use the SU(3)$_F$ isoscalar factors from Refs.~\cite{deSwart:1963pdg,Kaeding:1995vq}, along with
SU(2) Clebsch-Gordan coefficients, 
to construct products of SU(3)$_F$ states.

There are 16 $\btos$ and 16 $\btod$ charmless three-body $B\to PPP$ decays, where $P = \pi$ or $K$. Under SU(3)$_F$, all three final-state particles belong to the same multiplet (an octet of SU(3)$_F$), and hence they can be treated as identical, so the six possible permutations of these particles must be considered. The FA final state is antisymmetric under the exchange of any two final-state particles. This is only possible when all three final-state pseudoscalars are distinct, which reduces the number of available decays to 11 for each of $\btos$ and $\btod$\footnote{There is a caveat here: two $\btos$ and two $\btod$ decays contain $K^0 {\bar K}^0$ in the final state. These have an FA state only if this pair can be detected as $K_S K_L$.}.

For the FA final state, one wants to find $({\bf 8}\times{\bf 8}\times{\bf 8})_{\rm FA}$.
The decomposition for ${\bf 8} \times {\bf 8} = {\bf 27} + {\bf 10} + {\bf 10^*} + {\bf 8} + {\bf 8} + {\bf 1}$ can be separated into ${\bf 27} + {\bf 8} + {\bf 1}$ (total 36) symmetric and ${\bf 10} + {\bf 10^*} + {\bf 8}$ (total 28) antisymmetric under the exchange of the two ${\bf 8}$'s. The FA irreps of ${\bf 8} \times {\bf 8} \times {\bf 8}$ arise from the product of the antisymmetric ${\bf 10} + {\bf 10^*} + {\bf 8}$ with the remaining ${\bf 8}$. This yields an FA final state that has dimension 56
under SU(3)$_F$. It can be decomposed into irreps of SU(3)$_F$ as follows:
\bea
({\bf 8}\times{\bf 8}\times{\bf 8})_{\rm FA} ={\bf 27}_{\rm FA} + {\bf 10}_{\rm FA} + {\bf 10^*}_{\rm FA} + {\bf 8}_{\rm FA} + {\bf 1},~~
\eea
where
\bea
{\bf 27}_{\rm FA} &=& \frac{2}{3}{\bf 27}_{{\bf 10}\times{\bf 8}} - \frac{2}{3}{\bf 27}_{{\bf 10^*}\times{\bf 8}} + \frac{1}{3}{\bf 27}_{{\bf 8}\times{\bf 8}} ~, \nl
{\bf 10}_{\rm FA} &=& -\sqrt{\frac{2}{3}}{\bf 10}_{{\bf 10}\times{\bf 8}} + \sqrt{\frac{1}{3}}{\bf 10}_{{\bf 8}\times{\bf 8}} ~, \nl
{\bf 10^*}_{\rm FA} &=& -\sqrt{\frac{2}{3}}{\bf 10^*}_{{\bf 10^*}\times{\bf 8}} + \sqrt{\frac{1}{3}}{\bf 10^*}_{{\bf 8}\times{\bf 8}} ~, \nl
{\bf 8}_{\rm FA} &=& \frac{1}{\sqrt{6}}{\bf 8}_{{\bf 10}\times{\bf 8}} + \frac{1}{\sqrt{6}}{\bf 8}_{{\bf 10^*}\times{\bf 8}} + \sqrt{\frac{2}{3}}{\bf 8}_{{\bf 8}\times{\bf 8}} ~.
\eea

\subsection{\boldmath SU(3)$_F$ assignments of pseudoscalar mesons}

The light-quark states ($u$, $d$ and $s$) transform as the fundamental triplet (${\bf 3}$) of SU(3)$_F$. The antiquarks transform as the ${\bf 3^*}$ of SU(3)$_F$. The quarks and antiquarks can be assigned the following representations using the $\ll{\bf r} Y I I_3\>$ notation:
\bea
\ll u\> ~=~ \ll{\bf 3}\oth\hf\hf\>~,&~~~~~~&-\ll{\bar u}\> ~=~ \ll{\bf 3^*}{-}\oth\hf{-}\hf\>~, \nl
\ll d\> ~=~ \ll{\bf 3}\oth\hf{-}\hf\>~,&~~~~~~&\ll{\bar d}\> ~=~ \ll{\bf 3^*}{-}\oth\hf\hf\>~, \nl
\ll s\> ~=~ \ll{\bf 3}{-}\tth00\>~,&~~~~~~&\ll{\bar s}\> ~=~ \ll{\bf 3^*}\tth00\>~.
\eea

The pions, kaons, and the octet component of the eta meson ($\eta_8$) form an octet (${\bf 8}$) of SU(3)$_F$, while the $\eta_1$ is an SU(3)$_F$ singlet. The physical $\eta$ and $\eta'$ mesons are linear combinations of the $\eta_8$ and $\eta_1$, constructed through octet-singlet mixing. In this work, we avoid the complications arising from this mixing by limiting our analysis to final states with only pions and/or kaons. The three pions and the four kaons are as follows:
\begin{widetext}
\bea
\ll\pi^+\> ~=~ \ll u\>\ll{\bar d}\>~=~\ll{\bf 8}011\>~,~~~~~\ll\pi^-\> ~=~ -\ll d\>\ll{\bar u}\>~=~\ll{\bf 8}01{-}1\>~,~~~~~~~ \nl
\ll\pi^0\> ~=~ \frac{\ll d\>\ll{\bar d}\> - \ll u\>\ll{\bar u}\>}{\s}~=~\ll{\bf 8}010\>~,~~~~~~~~~~~~~~~~~~~~~~~~~~\nl
\ll K^+\> ~=~ \ll u\>\ll{\bar s}\>~=~\ll{\bf 8}1\hf\hf\>~,~~~~~\ll K^0\> ~=~ \ll d\>\ll{\bar s}\>~=~\ll{\bf 8}1\hf{-}\hf\>~,~~~~~~~ \nl
\ll\ok\> ~=~ \ll s\>\ll{\bar d}\>~=~\ll{\bf 8}{-}1\hf\hf\>~,~~~~~\ll K^-\> ~=~ -\ll s\>\ll{\bar u}\>~=~\ll{\bf 8}{-}1\hf{-}\hf\>~.
\eea
\end{widetext}

\subsection{\boldmath Fully-antisymmetric three-body final states}

We now construct the normalized FA $P_1P_2P_3$ final states within SU(3)$_F$. The FS final state studied in Ref.~\cite{Bhattacharya:2014eca} could be divided into three cases, depending on the number of truly identical particles in the final state. For the FA state, there is only one case: in order for the FA final state to be non-vanishing, all three final-state pseudoscalars must be distinct from one another (e.g., $\pi^0\pi^+\pi^-$). We first construct states that are antisymmetrized over the first two particles. We then add all three combinations antisymmetrized in this way
to obtain the FA state.

In what follows the state is antisymmetrized over particles that are included within square brackets:
\begin{widetext}
\bea
\label{eq:sym}
\ll [P_1 P_2] P_3 \> &=& \frac{1}{\s}\[\ll P_1\>\ll P_2\>\ll P_3\> - \ll P_2\>\ll P_1\>\ll P_3\>\] ~,\nl
\ll [P_1 P_2 P_3] \>_{\rm FA} &=& \frac{1}{\st}\[\ll [P_1 P_2] P_3\> + \ll [P_2 P_3] P_1\> + \ll [P_3 P_1] P_2\>\] ~.
\eea
\end{widetext}
Note that, if any two of three (or all three) of the particles are identical (e.g., $\pi^0\pi^0\pi^+$ or $\pi^0\pi^0\pi^0$), the three-particle state, $\ll [P_1 P_2 P_3] \>_{\rm FA}$, automatically vanishes.

\subsection{\boldmath Three-body $\btos$ and $\btod$ transitions using SU(3)$_F$}

The Hamiltonian for three-body $B$ decays follows from the underlying
quark-level transitions $\btos q{\bar q}$ and $\btod q{\bar q}$, where
$q$ is an up-type quark ($u, c, t$). However, the unitarity of the Cabibbo-Kobayashi-Maskawa (CKM)
matrix, given as
\bea
\sum\limits_{q = u, c, t}V^*_{qb}V^{}_{qs} = 0 ~~,~~~~
\sum\limits_{q = u, c, t}V^*_{qb}V^{}_{qd} = 0~,
\label{unitarity}
\eea
allows us to trade one of the up-type quarks for the other two.  Here
we choose to replace the $t$-quark operators and retain only the
$c$-quark and $u$-quark operators. Thus
the weak-interaction
Hamiltonian is composed of four types of operators: $\btos
c{\bar c}$, $\btod c{\bar c}$, $\btos u {\bar u}$, and $\btod u{\bar u}$.

The SU(3)$_F$ representations of these operators are
dictated by the light quarks since the heavy $b$, $c$, and $t$
quarks
are SU(3)$_F$ singlets. The transition operators are given as follows:
\begin{widetext}
\bea
{\cal O}_{\btos c{\bar c}} &=& V^{*}_{cb}V^{}_{cs}B^{(\bf 3^*)}_{\{\tth, 0, 0\}} ~, ~~~~~
{\cal O}_{\btod c{\bar c}} ~=~ V^{*}_{cb}V^{}_{cd}B^{(\bf 3^*)}_{\{-\oth, \hf, \hf\}}  ~, \nl
{\cal O}_{\btos u{\bar u}} &=& V^{*}_{ub}V^{}_{us}\lb A^{(\bf 3^*)}_{\{\tth, 0, 0\}} + R^{(\bf 6)}_{\{\tth, 1, 0\}} + \sx P^{(\bf 15^*)}_{\{\tth, 1, 0\}} + \st P^{(\bf 15^*)}_{\{\tth, 0, 0\}}\rb ~,\nl
{\cal O}_{\btod u{\bar u}} &=& V^{*}_{ub}V^{}_{ud}\lb A^{(\bf 3^*)}_{\{-\oth, \hf, \hf\}} - R^{(\bf 6)}_{\{-\oth, \hf, \hf\}} + \sqrt{8} P^{(\bf 15^*)}_{\{-\oth, \thf, \hf\}} + P^{(\bf 15^*)}_{\{-\oth, \hf, \hf\}}\rb ~,
\eea
\end{widetext}
where we have used the notation $O^{(\bf r)}_{\{Y, I, I_3\}}$
to represent each SU(3)$_F$ operator ($O = \{A, B, R, P\}$).
We have taken the names of these operators and their relative signs
from Ref.~\cite{Zeppenfeld:1980ex}. The weak-interaction Hamiltonian
that governs charmless $B$ decays is then simply the sum of these
four operators:
\bea
{\cal H} &=& {\cal O}_{\btos c{\bar c}} + {\cal O}_{\btod c{\bar c}} + {\cal O}_{\btos u{\bar u}} + {\cal O}_{\btod u{\bar u}}  ~.
\eea

The above Hamiltonian governs the decay of the SU(3)$_F$ triplet of $B$-mesons [$B^{\bf 3} = (B^+_u, B^0_d, B^0_s$)], whose components have the same SU(3)$_F$ representations as their corresponding light quarks. The fully-antisymmetric three-body decay amplitude for the process $B \to P_1 P_2 P_3$ can now be constructed easily as follows:
\bea
{\cal A}_{\rm FA}(p_1, p_2, p_3) &=& _{\rm FA}\<[P_1P_2P_3]\ll\rr{\cal H}\ll\rr B^{\bf 3}\>~,
\eea
where $p_i$ represents the momentum of the final-state particle $P_i$.

\begin{table*}[h]
\renewcommand*{\arraystretch}{1.9}
\caption{Amplitudes for $\Delta S = 1$ $B$-meson decays to
  fully-antisymmetric $PPP$ states as functions of nine SU(3)$_F$ RMEs.}
\label{tab:1}
\begin{center}
\begin{tabular}{|c|cc|ccccccc|} \hline \hline
Decay & \multicolumn{2}{c|}{$V^*_{cb}V^{}_{cs}$} & \multicolumn{7}{c|}{$V^*_{ub}V^{}_{us}$} \\ \cline{2-10}
Amplitude & $B^{\FA}_1$ & $B^{\FA}$ & $A^{\FA}_1$ & $A^{\FA}$ & $R^{\FA}_8$ & $R^{\FA}_{10}$ & $P^{\FA}_8$ & $P^{\FA}_{10^*}$ & $P^{\FA}_{27}$ \\ \hline \hline
$\s\cA(B^+\to K^+\pi^+\pi^-)_{\rm FA}$ & 0 & $\ssof$ & 0 & $\ssof$ & $\ssofi$ & $\ssot$ & $-\frac{3\s}{5}$ & 0 & $\frac{6\st}{5}$ \\
$\cA(B^+\to K^0\pi^+\pi^0)_{\rm FA}$ & 0 & $-\ssof$ & 0 & $-\ssof$ & $-\ssofi$ & $\frac{1}{\sx}$ & $\frac{3\s}{5}$ & 0 & $-\frac{\st}{5}$ \\
$\s\cA(\bd\to K^0\pi^+\pi^-)_{\rm FA}$ & 0 & $-\ssof$ & 0 & $-\ssof$ & $\ssofi$ & $\ssot$ & $-\frac{\s}{5}$ & 0 & $\frac{2\st}{5}$ \\
$\cA(\bd\to K^+\pi^0\pi^-)_{\rm FA}$ & 0 & $\ssof$ & 0 & $\ssof$ & $-\ssofi$ & $\frac{1}{\sx}$ & $\frac{\s}{5}$ & 0 & $\frac{3\st}{5}$ \\  \hline \hline
$\s\cA(B^+\to K^+K^0\ok)_{\rm FA}$ & 0 & $\ssof$ & 0 & $\ssof$ & $\ssofi$ & 0 & $-\frac{3\s}{5}$ & 0 & $-\frac{4\st}{5}$ \\
$\s\cA(\bd\to K^0K^+K^-)_{\rm FA}$ & 0 & $-\ssof$ & 0 & $-\ssof$ & $\ssofi$ & 0 & $-\frac{\s}{5}$ & $2$ & $\frac{2\st}{5}$ \\ \hline \hline
$\cA(\bs\to\pi^0K^+K^-)_{\rm FA}$ & $-\frac{1}{4\st}$ & 0 & $-\frac{1}{4\st}$ & 0 & $\ssofi$ & $-\frac{1}{2\sx}$ & $\frac{2\s}{5}$ & $-\frac{1}{2}$ & $-\frac{\st}{20}$ \\
$\cA(\bs\to\pi^0K^0\ok)_{\rm FA}$ & $-\frac{1}{4\st}$ & 0 & $-\frac{1}{4\st}$ & 0 & $-\ssofi$ & $\frac{1}{2\sx}$ & $-\frac{2\s}{5}$ & $\frac{1}{2}$ & $-\frac{9\st}{20}$ \\
$\s\cA(\bs\to\pi^-K^+\ok)_{\rm FA}$ & $\frac{1}{2\st}$ & 0 & $\frac{1}{2\st}$ & 0 & 0 & $\frac{1}{\sx}$ & 0 & $-1$ & $\frac{\st}{2}$ \\
$\s\cA(\bs\to\pi^+K^0K^-)_{\rm FA}$ & $\frac{1}{2\st}$ & 0 & $\frac{1}{2\st}$ & 0 & 0 & $-\frac{1}{\sx}$ & 0 & $1$ & $\frac{\st}{2}$ \\ \hline\hline
$\cA(\bs\to\pi^0\pi^+\pi^-)_{\rm FA}$ & $-\frac{1}{2\st}$ & $\ssof$ & $-\frac{1}{2\st}$ & $\ssof$ & 0 & 0 & $\frac{3\s}{5}$ & 0 & $\frac{3\st}{10}$ \\ \hline\hline
\end{tabular}
\end{center}
\end{table*}

\begin{table*}[h]
\renewcommand*{\arraystretch}{1.9}
\caption{Amplitudes for $\Delta S = 0$ $B$-meson decays to fully-antisymmetric $PPP$ states as functions of SU(3)$_F$ RMEs.} \label{tab:2}
\begin{center}
\begin{tabular}{|c|cc|ccccccc|} \hline \hline
Decay & \multicolumn{2}{c|}{$V^*_{cb}V^{}_{cd}$} & \multicolumn{7}{c|}{$V^*_{ub}V^{}_{ud}$} \\ \cline{2-10}
Amplitude & $B^{\FA}_1$ & $B^{\FA}$ & $A^{\FA}_1$ & $A^{\FA}$ & $R^{\FA}_8$ & $R^{\FA}_{10}$ & $P^{\FA}_8$ & $P^{\FA}_{10^*}$ & $P^{\FA}_{27}$ \\ \hline \hline
$\cA(B^+\to\pi^+ K^0\ok)_{\rm FA}$ & 0 & $-\oof$ & 0 & $-\oof$ & $-\oofi$ & 0 & $\frac{3}{5}$ & 0 & $\frac{2\sx}{5}$ \\
$\cA(B^+\to\pi^+K^+K^-)_{\rm FA}$ & 0 & $\oof$ & 0 & $\oof$ & $\oofi$ & $\oot$ & $-\frac{3}{5}$ & 0 & $\frac{3\sx}{5}$ \\
$\s\cA(B^+\to\pi^0K^+\ok)_{\rm FA}$ & 0 & 0 & 0 & 0 & 0 & $-\oot$ & 0 & 0 & $\sx$ \\
$\s\cA(\bd\to\pi^0K^0\ok)_{\rm FA}$ & $-\frac{1}{2\sx}$ & $-\oof$ & $-\frac{1}{2\sx}$ & $-\oof$ & $-\oofi$ & $-\frac{1}{2\st}$ & -1 & $-\frac{1}{\s}$ & $\frac{3\st}{2\s}$ \\
$\s\cA(\bd\to\pi^0K^+K^-)_{\rm FA}$ & $-\frac{1}{2\sx}$ & $\oof$ & $-\frac{1}{2\sx}$ & $\oof$ & $\oofi$ & $\frac{1}{2\st}$ & 1 & $\frac{1}{\s}$ & $\frac{3\st}{2\s}$  \\
$\cA(\bd\to\pi^+K^0K^-)_{\rm FA}$ & $\frac{1}{2\sx}$ & 0 & $\frac{1}{2\sx}$ & 0 & 0 & $\frac{1}{2\st}$ & 0 & $-\frac{1}{\s}$ & $\frac{\st}{2\s}$ \\
$\cA(\bd\to\pi^-K^+\ok)_{\rm FA}$ & $\frac{1}{2\sx}$ & 0 & $\frac{1}{2\sx}$ & 0 & 0 & $-\frac{1}{2\st}$ & 0 & $\frac{1}{\s}$ & $\frac{\st}{2\s}$ \\
 \hline \hline
$\s\cA(\bd\to\pi^0\pi^+\pi^-)_{\rm FA}$ & $-\frac{1}{\sx}$ & $-\oof$ & $-\frac{1}{\sx}$ & $-\oof$ & $\frac{\st}{\sf}$ & 0 & $\frac{3}{5}$ & 0 & $-\frac{\st}{5\s}$ \\
 \hline\hline
$\cA(\bs\to\ok\pi^+\pi^-)_{\rm FA}$ & 0 & $-\oof$ & 0 & $-\oof$ & $\oofi$ & 0 & $-\frac{1}{5}$ & $\s$ & $\frac{\sx}{5}$ \\
$\s\cA(\bs\to K^-\pi^+\pi^0)_{\rm FA}$ & 0 & $\frac{2}{\sf}$ & 0 & $\frac{2}{\sf}$ & $-\frac{2}{\sqrt{15}}$ & 0  & $\frac{2}{5}$ & $\s$ & $\frac{3\sx}{5}$ \\
 \hline \hline
$\cA(\bs\to\ok K^+K^-)_{\rm FA}$ & 0 & $-\oof$ & 0 & $-\oof$ & $\oofi$ & $\oot$ & $-\frac{1}{5}$ & 0 & $\frac{\sx}{5}$ \\ \hline\hline
\end{tabular}
\end{center}
\end{table*}

\subsection{Reduced matrix elements}

The 22 charmless three-body $B$ decay amplitudes (11 $\btos$ and 11 $\btod$) can all be written in terms of nine SU(3)$_F$ RMEs (the $Y, I$, and $I_3$ indices of the operators have been suppressed):
\bea
B^{\FA}_1 &\equiv& _{\rm FA}\<{\bf 1}\ll\ll B^{(\bf 3^*)}\rr\rr{\bf 3}\> ~, \nl
B^{\FA}   &\equiv& _{\rm FA}\<{\bf 8}\ll\ll B^{(\bf 3^*)}\rr\rr{\bf 3}\> ~, \nl
A^{\FA}_1 &\equiv& _{\rm FA}\<{\bf 1}\ll\ll A^{(\bf 3^*)}\rr\rr{\bf 3}\> ~, \nl
A^{\FA}   &\equiv& _{\rm FA}\<{\bf 8}\ll\ll A^{(\bf 3^*)}\rr\rr{\bf 3}\> ~, \nl
R_8^{\FA} &\equiv& _{\rm FA}\<{\bf 8}\ll\ll R^{(\bf 6)}\rr\rr{\bf 3}\> ~, \nl
R_{10}^{\FA} &\equiv& _{\rm FA}\<{\bf 10}\ll\ll R^{(\bf 6)}\rr\rr{\bf 3}\> ~, \nl
P_8^{\FA} &\equiv& _{\rm FA}\<{\bf 8}\ll\ll P^{(\bf 15^*)}\rr\rr{\bf 3}\> ~, \nl
P_{10^*}^{\FA} &\equiv& _{\rm FA}\<{\bf 10^*}\ll\ll P^{(\bf 15^*)}\rr\rr{\bf 3}\> ~, \nl
P_{27}^{\FA} &\equiv& _{\rm FA}\<{\bf 27}\ll\ll P^{(\bf 15^*)}\rr\rr{\bf 3}\> ~.
\eea

The decomposition of all 22 amplitudes in terms of these RMEs is given in Tables \ref{tab:1} and \ref{tab:2}. As in the FS case \cite{Bhattacharya:2014eca}, there are only seven combinations of matrix elements in the amplitudes since $B^{\FA}$ and $A^{\FA}$, as well as $B^{\FA}_1$ and
$A^{\FA}_1$, always appear together:
\bea
V^*_{cb} V^{}_{cq} \, B^{\FA} &+& V^*_{ub} V^{}_{uq} \, A^{\FA} ~,~ \nl
V^*_{cb} V^{}_{cq} \, B^{\FA}_1 &+& V^*_{ub} V^{}_{uq} \, A^{\FA}_1 ~.
\label{ABcombs}
\eea

\section{Diagrams}
\label{sec:diags}

In Refs.~\cite{Gronau:1994rj, Gronau:1995hn}, flavor-flow diagrams were proposed to describe two-body $B \to PP$ decays. There are eight diagrams: $T$ (tree), $C$ (color-suppressed tree), $P$ (penguin), $E$ (exchange), $A$ (annihilation), $PA$ (penguin annihilation), $P_{EW}$ (electroweak penguin [EWP]), and $P^C_{EW}$ (color-suppressed EWP)\footnote{Note that EWPs break isospin symmetry due to the difference in the $Z$-boson's couplings to up- and down-type quarks. However, in Eq.~(\ref{eq:redef}), we show that the EWPs can be absorbed into six other diagrams using redefinition rules. Therefore, the inclusion of EWPs does not impact our analysis of equivalence between diagrams and RMEs.}.

\subsection{\boldmath $B \to PPP$ decays}
\label{BPPPdiagrams}

Diagrams can also be used to describe $B \to PPP$ decays \cite{Lorier:2010xf, Bhattacharya:2014eca}. These closely follow those used in $B \to PP$ decays. For the three-body analogs of $T$, $C$, $P$, $P_{EW}$, and $P^C_{EW}$, one has to ``pop'' a quark pair from the vacuum.  The subscript ``1'' (``2'') is added if the popped quark pair is between two nonspectator final-state quarks (two final-state quarks including the spectator). One therefore has $T_i$, $C_i$, $P_{EWi}$, and $P^C_{EWi}$ diagrams, $i = 1,2$. It turns out that $P$-type diagrams only ever appear in amplitudes in the combination ${\tilde P} \equiv P_1 + P_2$. For the three-body analogs of $E$, $A$, and $PA$, the spectator quark interacts with the ${\bar b}$, and one has two popped quark pairs. Here there is only one of each type of diagram. Finally, for each of ${\tilde P}$ and $PA$, two contributions are allowed, namely ${\tilde P}_{ut}$, ${\tilde P}_{ct}$, $PA_{ut}$ and $PA_{ct}$, where ${\tilde P}_{ut} \equiv {\tilde P}_u - {\tilde P}_t$, and similarly for the other diagrams.

All diagrams involve products of CKM matrix elements. We define
\beq
\lambda_p^{(q)} \equiv V^*_{pb} V_{pq} ~,~~ q = d,s ~,~~ p = u, c, t ~.
\eeq
The diagrams $T_i$, $C_i$, ${\tilde P}_{ut}$, $E$, $A$ and $PA_{ut}$ all involve
$\lambda_u^{(q)}$, ${\tilde P}_{ct}$ and $PA_{ct}$ involve $\lambda_c^{(q)}$, and $P_{EWi}$, and $P^C_{EWi}$ involve $\lambda_t^{(q)}$. In this section, we use the convention in which the $\lambda_p^{(q)}$ factors are contained completely (magnitude and phase) in the diagrams\footnote{Attention: this convention is not universal for analyses with diagrams. Indeed, in Sec.~\ref{sec_Applications}, we use a different convention -- there only the magnitude of $\lambda_p^{(q)}$ is absorbed into the diagrams}.

The four EWP diagrams, $P_{EW1,2}$ and $P^C_{EW1,2}$, are not really independent: their addition only has the effect of redefining other diagrams. The following redefinition rules can be used to absorb the four EWP diagrams into six other diagrams:
\bea
T_1 & \to & T_1 + P^C_{EW1} ~, \nn \\
T_2 & \to & T_2 - P^C_{EW2} ~, \nn \\
C_1 & \to & C_1 + P_{EW1} ~, \nn \\
C_2 & \to & C_2 - P_{EW2} ~, \nn \\
({\tilde P}_{ut} + {\tilde P}_{ct}) & \to & ({\tilde P}_{ut} + {\tilde P}_{ct}) + \frac13 (P^C_{EW1} + P^C_{EW2}) ~\label{eq:redef}.
\eea
Note that, before redefinition, $T_i$, $C_i$, ${\tilde P}_{ut}$, and  ${\tilde P}_{ct}$ each involve only a single product of CKM matrix elements, $\lambda_p^{(q)}$. After redefinition, this is no longer true.

There are therefore in total ten diagrams, namely $T_{1,2}$, $C_{1,2}$, ${\tilde P}_{ct}$, ${\tilde P}_{ut}$, $E$, $A$, $PA_{ct}$, and $PA_{ut}$. The decomposition of all 22 amplitudes in terms of the diagrams is given in Tables \ref{tab:3} and \ref{tab:4}.

\begin{table*}
\renewcommand*{\arraystretch}{2}
\caption{Amplitudes for $\Delta S = 1$ $B$-meson decays to
  fully-antisymmetric $PPP$ states as a function of the three-body
  diagrams ($\btos$ diagrams are written with primes).}
\label{tab:3}
\begin{center}
\begin{tabular}{|c|cc|cccccccc|} \hline \hline
Decay & \multicolumn{2}{c|}{$V^*_{cb}V^{}_{cs}$} & \multicolumn{8}{c|}{$V^*_{ub}V^{}_{us}$} \\ \cline{2-11}
Amplitude & ${\tilde P}'_{ct}$ & $PA'_{ct}$ & ${\tilde P}'_{ut}$ & $PA'_{ut}$ & $C'_1$ & $C'_2$ & $T'_1$ & $T'_2$ & $E'$ & $A'$ \\ \hline \hline
$\s\cA(B^+ \to K^+\pi^+\pi^-)$ & $\s$  & 0 & $\s$  & 0 & $-\s$ & 0 & 0 & $\s$ & 0 & $-\s$ \\
$\cA(B^+ \to K^0\pi^+\pi^0)$    & $-\s$ & 0 & $-\s$ & 0 & 0 & $\ons$ & $\ons$ & 0 & 0 & $\s$ \\
$\s\cA(\bd \to K^0\pi^+\pi^-)$ & $-\s$ & 0 & $-\s$ & 0 & $-\s$ & 0 & $\s$ & 0 & 0 & 0 \\
$\cA(\bd \to K^+\pi^0\pi^-)$    & $\s$  & 0 & $\s$  & 0 & 0 & $\ons$ & $-\ons$ & $\s$ & 0 & 0 \\
\hline \hline
$\s\cA(B^+ \to K^+K^0\ok)$     & $\s$  & 0 & $\s$  & 0 & 0 & 0 & 0 & 0 & 0 & $-\s$ \\
$\s\cA(\bd \to K^0K^+K^-)$     & $-\s$ & 0 & $-\s$ & 0 & $-\s$ & 0 & 0 & $-\s$ & 0 & 0 \\
\hline \hline
$\cA(\bs \to \pi^0 K^+K^-)$ & $-\ons$ & $\ons$ & $-\ons$ & $\ons$ & 0 & $-\ons$ & $\ons$ & $-\ons$ & $\s$ & 0 \\
$\cA(\bs \to \pi^0 K^0\ok)$ & $-\ons$ & $\ons$ & $-\ons$ & $\ons$ & 0 & $\ons$ & 0 & 0 & 0 & 0 \\
$\s\cA(\bs \to \pi^- K^+\ok)$ & $\s$ & $-\s$ & $\s$ & $-\s$ & 0 & 0 & 0 & $\s$ & $-\s$ & 0 \\
$\s\cA(\bs \to \pi^+ K^0K^-)$ & $\s$ & $-\s$ & $\s$ & $-\s$ & 0 & 0 & $-\s$ & 0 & $-\s$ & 0 \\
\hline\hline
$\cA(\bs \to \pi^0\pi^+\pi^-)$ & 0 & $\s$ & 0 & $\s$ & 0 & 0 & 0 & 0 & $\ths$ & 0 \\
\hline\hline
\end{tabular}
\end{center}
\end{table*}

\begin{table*}
\renewcommand*{\arraystretch}{2}
\caption{Amplitudes for $\Delta S = 0$ $B$-meson decays to fully-antisymmetric $PPP$ states as a function of the three-body diagrams.}
\label{tab:4}
\begin{center}
\begin{tabular}{|c|cc|cccccccc|} \hline \hline
Decay & \multicolumn{2}{c|}{$V^*_{cb}V^{}_{cd}$} & \multicolumn{8}{c|}{$V^*_{ub}V^{}_{ud}$} \\ \cline{2-11}
Amplitude & ${\tilde P}_{ct}$ & $PA_{ct}$ & ${\tilde P}_{ut}$ & $PA_{ut}$ & $C_1$
& $C_2$ & $T_1$ & $T_2$ & $E$ & $A$ \\ \hline \hline
$\cA(B^+ \to \pi^+K^0\ok)$ & -1 & 0 & -1 & 0 & 0 & 0 & 0 & 0 & 0 & 1 \\
$\cA(B^+ \to \pi^+K^+K^-)$ & 1 & 0 & 1 & 0 & -1 & 0 & 0 & 1 & 0 & -1 \\
$\s\cA(B^+\to\pi^0K^+\ok)$ & 0 & 0 & 0 & 0 & 0 & -1 & -1 & 0 & 0 & 0 \\
$\s\cA(\bd \to \pi^0K^0\ok)$ & -2 & 1 & -2 & 1 & 0 & -1 & 0 & 0 & 0 & 0 \\
$\s\cA(\bd \to \pi^0K^+K^-)$ & 0 & 1 & 0 & 1 & -1 & 0 & 0 & 0 & 2 & 0 \\
$\cA(\bd \to \pi^+K^0K^-)$ & 1 & -1 & 1 & -1 & 0 & 0 & 0 & 1 & -1 & 0 \\
$\cA(\bd \to \pi^-K^+\ok)$ & 1 & -1 & 1 & -1 & 0 & 0 & -1 & 0 & -1 & 0 \\
\hline \hline
$\s\cA(\bd \to \pi^0\pi^+\pi^-)$ & -3 & 2 & -3 & 2 & -1 & -1 & 2 & -2 & 3 & 0 \\
\hline\hline
$\cA(\bs\to \ok\pi^+\pi^-)$ & -1 & 0 & -1 & 0 & -1 & 0 & 0 & -1 & 0 & 0 \\
$\s\cA(\bs \to K^-\pi^+\pi^0)$ & 2 & 0 & 2 & 0 & 0 & 1 & -2 & 1 & 0 & 0 \\
\hline \hline
$\cA(\bs \to \ok K^+K^-)$ & -1 & 0 & -1 & 0 & -1 & 0 & 1 & 0 & 0 & 0 \\
\hline\hline
\end{tabular}
\end{center}
\end{table*}

\subsection{Equivalence of RMEs and diagrams}

In Sec.~\ref{sec:RMEs}, it was shown that the 11 $\btod$ decay amplitudes can be expressed in terms of nine RMEs, two of which contain $V_{cb}^*V_{cd}$ while seven others contain $V_{ub}^*V_{ud}$. In the previous subsection, we have seen that the same 11 $\btod$ decay amplitudes can also be expressed in terms of ten diagrams. By comparing the expressions for the amplitudes of the 11 $\btod$ decays, it is possible to express the nine RMEs in terms of the ten diagrams. These expressions are
\begin{widetext}
\bea
V_{cb}^*V_{cd}B_1^{\FA} &=& 2\sqrt{6}\({\tilde P}_{ct}-PA_{ct}\) ~, \\
V_{cb}^*V_{cd}B^{\FA} &=& \sqrt{5}P_{ct} ~, \\
V_{ub}^*V_{ud} A_1^{\FA} &=& \frac{\st}{2\s}\(8{\tilde P}_{ut} - 8PA_{ut} - 3T_1 + 3T_2 + C_1 + C_2 - 8E\)\label{eq:me2d1} ~, \\
V_{ub}^*V_{ud} A^{\FA} &=& \frac{\sf}{8}\(8{\tilde P}_{ut} - 3T_1 + 3T_2 + C_1 + C_2 + E - 3A\) ~, \\
V_{ub}^*V_{ud} R_8^{\FA} &=& \frac{\sqrt{15}}{4} \(T_1 - T_2 - C_1 - C_2 + E - A\) ~, \\
V_{ub}^*V_{ud}R_{10}^{\FA} &=& \frac{\st}{2} \(T_1 + T_2 - C_1 + C_2\) ~, \\
V_{ub}^*V_{ud}P_8^{\FA} &=& \frac{1}{8} \(T_1 - T_2 + C_1 + C_2 + 5E + 5A \) ~, \\
V_{ub}^*V_{ud}P_{10^*}^{\FA} &=& -\frac{1}{2\sqrt{2}} \(T_1 + T_2 + C_1 - C_2\) ~, \\
V_{ub}^*V_{ud}P_{27}^{\FA} &=& -\frac{1}{2\sx} \(T_1 - T_2 + C_1 + C_2\) ~.
\label{PPPrels}
\eea
\end{widetext}
Note that, in all of these relations, it naively appears that both sides involve the same products of CKM matrix elements. However, this is not really true -- as noted above, once the EWP contributions have been removed by redefining the other diagrams, these other diagrams no longer involve a well-defined product of CKM matrix elements.

By analyzing the 11 $\btos$ decays, one can similarly establish a corresponding set of expressions relating the diagrams to the RMEs for $\btos$ decays.

This demonstrates the equivalence of diagrams and SU(3)$_F$ for the fully-antisymmetric $PPP$ state.

\section{Amplitude Relations}
\label{sec:amprels}

Since all 22 decay amplitudes can be expressed in terms of seven combinations of RMEs, the amplitudes must obey 15 independent relationships in the SU(3)$_F$ limit. These relationships can be found as follows. The 11 $\btos$ decay amplitudes can be expressed in terms of the seven combinations of RMEs -- there must be four relations among these amplitudes. A subset of these relations can be obtained by considering processes related by isospin symmetry, while the remaining can be found using the full SU(3)$_F$ symmetry. The process can be repeated for the 11 $\btod$ decays generating four additional amplitude relations. The remaining seven relations follow from the application of U-spin symmetry that relates $\btos$ decays to $\btod$ decays.

\subsection{\boldmath $\btos$ Decays}

The 11 $\btos$ decays (see Table \ref{tab:1}) include four $B\to K\pi\pi$ decays, two $B\to KK{\bar K}$ decays, four $\bs\to \pi K{\bar K}$ decays, and one $\bs\to \pi\pi\pi$ decay. Each decay amplitude can be expressed as a linear combination of seven RMEs. Therefore, these amplitudes must satisfy four relationships. We find that the four $B\to K\pi\pi$ decays and the four $\bs\to\pi K{\bar K}$ decays each satisfy one quadrangle relationship, while two additional quadrangle relationships span multiple types of decays. These relations are:
\begin{widetext}
\begin{enumerate}
    \item $B\to K\pi\pi$:
        \bea
           \s\cA(B^+\to K^+\pi^+\pi^-)_{\rm FA} + \cA(B^+\to K^0\pi^+\pi^0)_{\rm FA} &=&
           \s\cA(\bd\to K^0\pi^+\pi^-)_{\rm FA} + \cA(\bd\to K^+\pi^0\pi^-)_{\rm FA}.
        \eea
    \item $\bs\to\pi K{\bar K}$:
        \bea
            \s\cA(\bs\to\pi^0K^+K^-)_{\rm FA} + \s\cA(\bs\to\pi^0K^0\ok)_{\rm FA} &=& 
            -~\cA(\bs\to\pi^-K^+\ok)_{\rm FA} - \cA(\bs\to\pi^+K^0K^-)_{\rm FA}.
        \eea
    \item $\bd\to K\pi\pi, \bs\to\pi K{\bar K}$, and $\bd\to KK{\bar K}$ or $\bs\to\pi\pi\pi$:
        \bea
            \cA(\bd\to K^0\pi^+\pi^-)_{\rm FA} - \cA(\bd\to K^0K^+K^-)_{\rm FA} &=&
            \cA(\bs\to\pi^-K^+\ok)_{\rm FA} - \cA(\bs\to\pi^+K^0K^-)_{\rm FA} ~, \label{eq:usb2s} \\
            \s\cA(\bd\to K^+\pi^0\pi^-)_{\rm FA} + \s\cA(\bs\to\pi^0K^+K^-)_{\rm FA} &=&
            \cA(\bs\to\pi^-K^+\ok)_{\rm FA} + \s\cA(\bs\to\pi^0\pi^+\pi^-)_{\rm FA} ~. \label{eq:3pi}
        \eea
\end{enumerate}
\end{widetext}

\subsection{\boldmath $\btod$ Decays}

The 11 $\btod$ decays (see Table \ref{tab:2}) include seven $B\to \pi K{\bar K}$ decays, one $\bd\to \pi\pi\pi$ decay, two $\bs\to K\pi\pi$ decays, and one $\bs\to {\bar K}K{\bar K}$ decay. Again, each decay amplitude can be expressed as a linear combination of seven RMEs, so that there must be four amplitude relationships. We find two quadrangle relationships among these amplitudes:
\begin{widetext}
\bea
\cA(\bd\to\pi^+K^0K^-)_{\rm FA} - \cA(\bd\to\pi^-K^+\ok)_{\rm FA} &=& \cA(\bs\to\ok K^+K^-)_{\rm FA} - \cA(\bs\to\ok\pi^+\pi^-)_{\rm FA} ~, \label{eq:usb2d} \\
\s\cA(\bd\to\pi^0K^+K^-)_{\rm FA} - \cA(\bd\to\pi^+K^0K^-)_{\rm FA} &=& \s\cA(\bd\to\pi^0\pi^+\pi^-)_{\rm FA} + \s\cA(\bs\to K^-\pi^+\pi^0)_{\rm FA} ~.
\eea
\end{widetext}
In addition, all seven $B\to\pi K{\bar K}$ decays satisfy one amplitude relationship, while another relationship involves multiple different amplitudes. As these relationships are not particularly enlightening, we do not present them here.

\subsection{U Spin}
\label{Sec:Uspin}

The final states in six $\btos$ decays in Table \ref{tab:1} and six corresponding $\btod$ decays in Table \ref{tab:2} do not involve any $\pi^0$s. Each pair of corresponding $\btos$ and $\btod$ decays is related by U-spin reflection ($d\leftrightarrow s$). The
six pairs are:
\begin{enumerate}

\item $\bd\to K^0\pi^+\pi^-$ and $\bs\to\ok K^+K^-$,
\item $\bd\to K^0K^+K^-$ and $\bs\to\ok\pi^+\pi^-$,
\item $\bs\to\pi^-K^+\ok$ and $\bd\to\pi^+K^0K^-$,
\item $\bs\to\pi^+K^0K^-$ and $\bd\to\pi^-K^+\ok$,	
\item $B^+\to K^+\pi^+\pi^-$ and $B^+\to\pi^+K^+K^-$,
\item $B^+\to K^+K^0\ok$ and $B^+\to\pi^+K^0\ok$,
	
\end{enumerate}
where the first (second) decay is $\btos$ ($\btod$). In each pair, amplitude terms multiplying $V^*_{cb}V^{}_{cs}$ and $V^*_{ub}V^{}_{us}$ in the $\btos$ process equal amplitude terms multiplying $V^*_{cb}V^{}_{cd}$ and $V^*_{ub}V^{}_{ud}$ in the $\btod$ process (up to an overall negative sign arising from the order of final-state particles \cite{Gronau:2000zy}; see Tables \ref{tab:1} and \ref{tab:2}). Thus, one can write relations among the $\btos$ and $\btod$ decay amplitudes
involving CKM matrix elements.

However, there is another relationship between U-spin pairs that is more useful experimentally \cite{Gronau:2003ep,Gronau:2000zy,Imbeault:2011jz}. It is
\beq
\frac{A_s}{A_d} \, \frac{{\cal B}_s}{{\cal B}_d} = -1 ~,
\label{Uspinobs}
\eeq
where
\bea
{\cal B}_d &=& |\cA(\bar b \to \bar d)|^2+|\cA(b \to d)|^2 ~,\nn\\
{\cal B}_s &=& |\cA(\bar b \to \bar s)|^2+|\cA(b \to s)|^2 ~,\nn\\
A_d &=& \frac{|\cA(\bar b \to \bar d)|^2-|\cA(b \to d)|^2}{|\cA(\bar b \to \bar d)|^2+|\cA(b \to d)|^2}~,\nn\\
A_s &=& \frac{|\cA(\bar b \to \bar s)|^2-|\cA(b \to s)|^2}{|\cA(\bar b \to \bar s)|^2+|\cA(b \to s)|^2}~.
\eea
${\cal B}_d$ and ${\cal B}_s$ are related to the CP-averaged $\btod$ and $\btos$ decay rates, while $A_d$ and $A_s$ are direct CP asymmetries. The CP-conjugate amplitude ${\bar\cA}(\bar b \to \bar q)$ is obtained from $\cA(b \to q)$ by changing the signs of the weak phases. These relations hold for all final symmetry states for all U-spin reflections.

There are six U-spin relations of this kind. Two additional U-spin amplitude relations connect several $\btos$ and $\btod$ decays. Since these additional relations are of no particular interest, we do not present them here. Along with the four $\btos$ and four $\btod$ decay amplitude relations, of which one pair [Eqs.~(\ref{eq:usb2s}) and (\ref{eq:usb2d})] is related by U-spin reflection, this makes a total of 15 independent relations.  This is consistent with the fact that 22 decay amplitudes are all expressed as a function of seven combinations of SU(3)$_F$ matrix elements.

\section{\boldmath Neglect of $E$/$A$/$PA$}
\label{Neglect E/A/PA}

We have seen in the previous sections that FA $B \to PPP$ decays can be written in terms of SU(3)$_F$ RMEs or in terms of diagrams, and that these descriptions are equivalent. However, the diagrammatic description does provide an additional useful tool.

When diagrams were introduced to describe $B \to PP$ amplitudes \cite{Gronau:1994rj, Gronau:1995hn}, it was noted that the description in terms of diagrams provides dynamical input. In particular, the diagrams $E$, $A$, and $PA$ all involve the interaction of the spectator quark. As such, they are expected to be considerably smaller than the $T$, $C$, and $P$ diagrams, and can therefore be neglected, to a first approximation. This reduces the number of unknown parameters and simplifies the analysis considerably. It must be stressed that this does not follow from group theory -- it is dynamical theoretical input. Even so, experimental measurements are consistent with this approximation; the branching ratios of processes that proceed only through $E$/$A$/$PA$ are indeed considerably smaller than those that are described by $T$/$P$/$C$.

With this in mind, it is likely that the $E$/$A$/$PA$ diagrams can be neglected in $B\to PPP$ decays. The neglect of these diagrams leads to relationships among the SU(3)$_F$ RMEs:
\bea
B_1^{\FA} &=& \frac{2\sx}{\sf} B^{\FA}~, \nn \\
A_1^{\FA} &=& \frac{2\sx}{\sf} A^{\FA}~, \nn \\
P_{8}^{\FA} &=& - \frac{\st}{2\s} P_{27}^{\FA}~.
\eea
The above relations reduce the number of combinations of RMEs in SU(3)$_F$. Because $B_1^{\FA}$ and $B^{\FA}$ always appear with $A^{\FA}_1$ and $A^{\FA}$, respectively [Eq.~(\ref{ABcombs})], the first and second relations only lead to a reduction of the number of RMEs by 1. An additional reduction by one RME can be attributed to the third relation. The total number of RMEs upon neglecting $E$/$A$/$PA$ diagrams is then 5, down from the original 7.

This leads to two additional relations among the $\btos$ amplitudes, and similarly for the $\btod$ amplitudes. For $\btos$, the additional relations are:
\begin{enumerate}

\item $\cA(\bs\to\pi^0\pi^+\pi^-)_{\rm FA} = 0$, i.e., the
decay $\bs\to\pi^0\pi^+\pi^-$ is pure $E'/A'/PA'$. This simplifies Eq.~(\ref{eq:3pi}) into a triangle relationship.

\item $\cA(B^+\to K^+\pi^+\pi^-)_{\rm FA} = \cA(\bd\to K^0K^+K^-)_{\rm FA} + 2\cA(\bs\to\pi^-K^+\ok)_{\rm FA}$.

\end{enumerate}
For $\btod$, they are:
\begin{enumerate}

\item $\cA(B^+\to\pi^+K^+K^-)_{\rm FA} = cA(\bd\to\pi^+K^0K^-)_{\rm FA}$ \\$ + \s\cA(\bd\to\pi^0K^+K^-)_{\rm FA}$,

\item $\cA(\bd\to\pi^+K^0K^-)_{\rm FA} + \cA(\bs\to\ok\pi^+\pi^-)_{\rm FA}$ \\ $ ~~~=~\s\cA(\bd\to\pi^0K^+K^-)_{\rm FA}$

\end{enumerate}

\section{Applications}
\label{sec_Applications}

In Sec.~\ref{sec:diags}, we established a one-to-one correspondence between SU(3)$_F$ RMEs and flavor-flow diagrams for the fully-antisymmetric $PPP$ state. By expressing all 22 $\btos$ and $\btod$ decay amplitudes in terms of both RMEs and diagrams, we showed that these approaches are equivalent. In this section, we go beyond the demonstration of this equivalence and explore predictions that can be tested experimentally.

\subsection{\boldmath Observing decays to the $PPP$ fully-antisymmetric state}

To obtain the fully-antisymmetric final state for a given $B \to PPP$ decay, one proceeds as follows \cite{Lorier:2010xf}. For the decay $B \to P_1 P_2 P_3$, one defines the three Mandelstam variables $s_{ij} \equiv \left( p_i + p_j \right)^2$, where $p_i$ is the momentum of each $P_i$. Only two of these three are independent. Say the $B\to P_1 P_2 P_3$ Dalitz plot is given in terms of $s_{12}$ and $s_{13}$. One can obtain the decay amplitude ${\cal M} (s_{12}, s_{13})$ describing this Dalitz plot by performing an isobar analysis. Here the amplitude is expressed as the sum of a non-resonant and several intermediate resonant contributions:
\beq
{\cal M} (s_{12}, s_{13}) = {\cal N}_{\rm DP}\sum\limits_j c_j e^{i \theta_j} F_j
(s_{12}, s_{13})~,
\eeq
where the index $j$ runs over all contributions. Each contribution is expressed in terms of isobar coefficients $c_j$ (magnitude) and $\theta_j$ (phase), and a dynamical wave function $F_j$. ${\cal N}_{\rm DP}$ is a normalization constant. The $F_j$ take different forms depending on the contribution. The $c_j$ and $\theta_j$ are extracted from a fit to the Dalitz-plot event distribution. With ${\cal M} (s_{12}, s_{13})$ in hand, one can construct the fully-antisymmetric amplitude. It is given simply by
\begin{widetext}
\bea \label{fulsym}
{\cal M}_{\rm FA}(s_{12},s_{13}) &=&
\frac{1}{\sqrt{6}} \left[ {\cal M}(s_{12},s_{13}) - {\cal M}(s_{13},s_{12})
- {\cal M}(s_{12},s_{23}) \right. \nn\\
&& \hskip2truecm \left.+~{\cal M}(s_{23},s_{12}) - {\cal M}(s_{23},s_{13})
+ {\cal M}(s_{13},s_{23}) \right]~,
\eea
\end{widetext}
where one uses the relationship $s_{12} + s_{13} + s_{23} = m_B^2 + m_{P_1}^2 + m_{P_2}^2 + m_{P_3}^2$ to express the third Mandelstam variable in terms of the first two. For any three-body decay for which a Dalitz plot has been measured, one can extract the fully-antisymmetric amplitude in the above fashion. The Dalitz plane can be divided into six regions by three lines of symmetry; along each line of symmetry, there is a pair of Mandelstam variables that are equal. It is sufficient to construct ${\cal M}_{\rm FA}$ in only one of these six regions as the other five regions do not contain additional information due to the fully-antisymmetric nature of ${\cal M}_{\rm FA}(s_{12},s_{13})$. In a similar vein, one can construct the fully-antisymmetric amplitude for the CP-conjugate process, $\overline{{\cal M}}_{\rm FA}$ from its measured Dalitz plot.

The fully-antisymmetric amplitudes for the process and its CP conjugate are not directly observable as these contain unknown phases. However, one can construct the following three linearly independent observables using these amplitudes:
\bea
{\cal X}_{\rm FA}(s_{12},s_{13}) &=& |{\cal M}_{\rm FA}(s_{12},s_{13})|^2 + |\overline{\cal M}_{\rm FA}(s_{12},s_{13})|^2 ,~ \nn\\
{\cal Y}_{\rm FA}(s_{12},s_{13}) &=& |{\cal M}_{\rm FA}(s_{12},s_{13})|^2 - |\overline{\cal M}_{\rm FA}(s_{12},s_{13})|^2 ,~ \nn\\
{\cal Z}_{\rm FA}(s_{12},s_{13}) &=& {\rm Im}\[{\cal M}_{\rm FA}^*(s_{12},s_{13})\overline{\cal M}_{\rm FA}(s_{12},s_{13})\] .~
\label{Dalitzplotobs}
\eea
For any given decay, the observables ${\cal X}_{\rm FA}, {\cal Y}_{\rm FA},$ and ${\cal Z}_{\rm FA}$ depend on the position in the Dalitz plot and are related to the CP-averaged decay rate, the direct CP asymmetry, and the indirect CP asymmetry. While ${\cal X}$ and ${\cal Y}$ exist for any three-body decay with three distinct particles in the final state, ${\cal Z}$ is a meaningful physical observable only for decays in which the final state is flavor neutral, such as $K^0K^+K^-$ or $\pi^0K^0\ok$.

\subsection{Confronting data}

As we saw in the previous section, when the $E$/$A$/$PA$ diagrams are neglected, the amplitudes can be written as functions of five combinations of RMEs. Four are proportional to $V^*_{ub}V_{uq}$, and the fifth is a linear combination of pieces proportional to $V^*_{ub}V_{uq}$ and $V^*_{cb}V_{cq}$. However, when it comes to using this parametrization to describe actual data, this counting must be reexamined. This is because some observables measure CP violation, which is sensitive to the weak phases of the CKM matrix. The RMEs proportional to $V^*_{ub}V_{uq}$ and $V^*_{cb}V_{cq}$ do not contribute equally to these observables. So, the number of RMEs that can be probed by the data is actually six, five proportional to $V^*_{ub}V_{uq}$ and one proportional to $V^*_{cb}V_{cq}$.

Turning to diagrams, the first thing is that we cannot redefine diagrams to absorb the EWPs, since that mixes pieces involving different CKM factors. Instead, we do the counting as follows. When $E$/$A$/$PA$ are neglected, there are ten diagrams. $T_i$, $C_i$, and ${\tilde P}_{ut}$ involve $\lambda_u^{(q)}$, ${\tilde P}_{ct}$ involves $\lambda_c^{(q)}$, and the four EWP diagrams involve $\lambda_t^{(q)}$. Here, it is important to use a different convention for the diagrams than that used in Sec.~\ref{BPPPdiagrams}. Here, the diagrams contain only the magnitudes of the $\lambda_p^{(q)}$; the phase information, including minus signs, is explicitly written as a factor multiplying the diagrams. The key point now is that, just as was the case in $B \to PP$ decays \cite{Neubert:1998pt, Neubert:1998jq, Gronau:1998fn}, the EWP diagrams are related to the tree diagrams. Taking the ratios of Wilson coefficients $c_1/c_2 = c_9/c_{10}$, which holds to about 5\%, the simplified form of these relations is
\begin{equation}
P_{EWi} = \kappa T_i ~~,~~~~ P^C_{EWi} = \kappa C_i ~,
\end{equation}
where
\begin{equation}
\kappa \equiv - \frac{3}{2} \frac{|\lambda_t^{(q)}|}{|\lambda_u^{(q)}|} \frac{c_9 + c_{10}}{c_1 + c_2} ~.
\end{equation}
These are the same EWP-tree relations as hold for the FS state; see Ref.~\cite{Imbeault:2010xg}. With this, there are six independent diagrams, of which two -- ${\tilde P}_{ct}$ and ${\tilde P}_{ut}$ -- always appear together as a linear combination.

\subsection{\boldmath Extracting $\gamma$}

\subsubsection{$B \to K\pi\pi$ and $B \to KK{\bar K}$ decays}

The method proposed in Ref.~\cite{Rey-LeLorier:2011ltd} and carried out in Refs.~\cite{Bhattacharya:2013cla, Bertholet:2018tmx} uses the FS states of three $B \to K\pi\pi$ and two $B \to KK{\bar K}$ decays. They are $B^0 \to K^+ \pi^0 \pi^-$, $B^0 \to K^0 \pi^+ \pi^-$, $B^+ \to K^+ \pi^+ \pi^-$, $B^0 \to K^0 K^+ K^-$, and $B^0 \to K^0 K^0 {\bar K}^0$ (with both $K^0$ and ${\bar K}^0$ identified as $K_S$). These are chosen because the amplitudes can be expressed as functions of only five combinations of diagrams (and not six).

However, this method cannot be applied to the FA states, since there is no such state for $B^0 \to K^0 K^0 {\bar K}^0$. Of the six $B \to K\pi\pi$ and $B \to KK{\bar K}$ decays listed in Table \ref{tab:1}, two are not used in the above method: $B^+\to K^0\pi^+\pi^0$ and $B^+ \to K^+ K^0 {\bar K}^0$. While the first decay clearly has an FA state, the second decay has one only if the $K^0 {\bar K}^0$ in the final state is detected as $K_S K_L$. While this may be possible experimentally, it is not easy, so we will not include this decay. 

In this case, the amplitudes for the five decays $B^0 \to K^+ \pi^0 \pi^-$, $B^0 \to K^0 \pi^+ \pi^-$, $B^+ \to K^+ \pi^+ \pi^-$, $B^0 \to K^0 K^+ K^-$, and $B^+\to K^0\pi^+\pi^0$ are functions of six diagrams, so there are 12 unknown theoretical parameters: six magnitudes of diagrams, five relative strong phases, and $\gamma$. And there are a total of 12 observables: the CP-averaged decay rates (${\cal X}_{\rm FA}$) and direct CP asymmetries (${\cal Y}_{\rm FA}$) for the five decays, and the indirect CP asymmetries (${\cal Z}_{\rm FA}$) of $B^0 \to K^0\pi^+\pi^-$ and $B^0 \to K^0 K^+ K^-$. With an equal number of observables and unknown theoretical parameters, $\gamma$ can be extracted from a fit, albeit with discrete ambiguities.

Now, it is expected that $|{\tilde P}_{uc}| \simeq \lambda^2 |{\tilde P}_{tc}|$, where $\lambda \equiv \sin\theta_C \simeq 0.22$, so it is not a bad approximation to neglect ${\tilde P}'_{uc}$. If one does this, there are now ten unknown theoretical parameters, which will reduce the discrete ambiguity in the extraction of $\gamma$. (In this case, it is possible to add a theoretical parameter parametrizing the breaking of SU(3)$_F$; see discussion below.)

\subsubsection{General analysis}

To date, methods to extract $\gamma$ from $B \to PPP$ decays have focused mainly on $\Delta S=1$ $B \to K\pi\pi$ and $B \to KK{\bar K}$ decays. However, there are many more decays, including $\Delta S=0$ processes and/or $\Delta S=1$ $B_s^0$ decays. Looking at Tables \ref{tab:1} and \ref{tab:2}, and eliminating those that (i) contain $K^0 {\bar K}^0$ in the final state and (ii) vanish when $E$/$A$/$PA$ are neglected, we see there are a total of 17 $B\to PPP$ decays that have an FA final state. All of these are functions of the same six diagrams, so there are in total 12 unknown theoretical parameters. If/when the Dalitz plots of these decays are measured, we have the potential to perform a fit to the data with many more observables than unknown parameters. (Of course, this also holds for the FS final state.) We will probably be able to extract $\gamma$ with no discrete ambiguity.

\subsection{\boldmath SU(3)$_F$ breaking}

In this entire discussion, it has been assumed that SU(3)$_F$ is a good symmetry. However, we know that SU(3)$_F$ is in fact broken, and these breaking effects will inevitably affect the extraction of $\gamma$. In some cases, it is possible to include new theoretical parameters in the fit that measure the size of SU(3)$_F$ breaking. Now, the fits are performed at a specific point in the Dalitz plot. But there is evidence that, when one averages over the entire Dalitz plot, the size of SU(3)$_F$ breaking is significantly reduced.

As described above, the amplitudes of the FS states of the three $B \to K\pi\pi$ and two $B \to KK{\bar K}$ decays used in the analysis of Ref.~\cite{Bertholet:2018tmx} are functions of five effective diagrams. As such there are ten unknown parameters. But there are 12 observables. In light of this, the $B\to K K{\bar K}$ amplitudes were multiplied by an additional SU(3)$_F$-breaking parameter $\alpha_{SU(3)}$. It represented the fact that, for these decays, one must pop an $s{\bar s}$ pair from the vacuum, while in $B \to K\pi\pi$ decays, a $u{\bar u}$ or $d{\bar d}$ pair is popped. In Ref.~\cite{Bertholet:2018tmx}, it was found that, while the value of the magnitude of $\alpha_{SU(3)}$ could be sizeable at a given point in the Dalitz plot, it could also have either sign. When averaged over the entire Dalitz plot, it was found that the effect of SU(3)$_F$ breaking was only at the percent level.

A similar technique can be used for the FA $B \to PPP$ states. The number and type of SU(3)$_F$-breaking parameters that are added to the amplitudes depend on how many more observables there are than unknown theoretical parameters. But in principle, it should be possible to add such parameters and see if, as was the case above, the size of SU(3)$_F$ breaking is actually reduced when averaged over the entire Dalitz plot.

Another technique for testing U-spin breaking by averaging over the Dalitz plot was discussed in Ref.~\cite{Bhattacharya:2014eca} for the fully-symmetric final state. This technique is to apply Eq.~(\ref{Uspinobs}) to two decays that are U-spin reflections of each other, in the presence of U-spin breaking. In terms of the Dalitz plot observables of Eq.~(\ref{Dalitzplotobs}), Eq.~(\ref{Uspinobs}) can be rewritten as
\bea
-\frac{{\cal Y}_{\rm FA}(\btos)}{{\cal Y}_{\rm FA}(\btod)} &=& Y_{\rm FA} ,~
\eea
where $Y_{\rm FA}$ is a real number that captures the amount of U-spin breaking. Under perfect U-spin symmetry $Y_{\rm FA} = 1$, however, its measured value may be $Y_{\rm FA} > 1$ or $Y_{\rm FA} < 1$ depending on the Dalitz plot point. Averaging over the Dalitz plot one can then test the amount of U-spin breaking in these decays. This technique can be applied to test U-spin breaking in the six U-spin-related pairs of decays listed in Sec.~\ref{Sec:Uspin}.

\section{Conclusions}

Recently, the CP-phase $\gamma$ was extracted from observables associated with the Dalitz plots of $\bd \to K^+\pi^0\pi^-$, $\bd \to K^0\pi^+\pi^-$, $B^+ \to K^+\pi^+\pi^-$, $\bd \to K^+ K^0 K^-$, and $\bd \to K^0 K^0 \ok$ \cite{Bertholet:2018tmx}. These decays all receive significant loop-level gluonic and/or electroweak penguin contributions, and so could be affected by NP. The presence of this NP would be revealed by a difference between the (loop-level) value of $\gamma$ found here and the value found using a standard method involving only tree-level decays \cite{Gronau:1990ra, Gronau:1991dp, Atwood:1996ci, Giri:2003ty}.

In three-body charmless $B \to PPP$ decays, there are six possibilities for the final state: a fully symmetric state, a fully antisymmetric state, or one of four mixed states. The analysis of Ref.~\cite{Bertholet:2018tmx} used the FS state and found six possible values for $\gamma$. One value agrees with that measured independently using tree-level decays, while the other five are in disagreement and hint at the presence of NP. In order to determine which of these is the true value of $\gamma$ in this system, one must extract $\gamma$ from a second set of $B \to PPP$ decays, this time using a different symmetry of the final state. There may again be multiple solutions, but the true value of $\gamma$ will be common to both analyses.

In this paper, we present the formalism describing charmless $B \to PPP$ decay amplitudes in which the final-state particles are all $\pi$'s or $K$'s, and the final state is fully antisymmetric. This can be used to perform analyses for extracting $\gamma$. In FA states, there are no identical particles in the final state; there are 11 $\btos$ and 11 $\btod$ $B \to PPP$ decays of this type. (But note that four decays have $K^0 {\bar K}^0$ in the final state. These have an FA state only if this pair can be detected as $K_S K_L$.) We write all 22 amplitudes in terms of seven combinations of nine SU(3)$_F$ reduced matrix elements. We also present the 15 relations among the amplitudes, some of which can be tested experimentally.

The amplitudes can also be written in terms of eight combinations of ten diagrams. By comparing the expressions for the amplitudes in terms of RMEs and diagrams, we are able to write the RMEs as functions of diagrams. This demonstrates the equivalence of diagrams and SU(3)$_F$. Diagrams also provide dynamical input: the three diagrams $E$, $A$, and $PA$ all involve the interaction of the spectator quark and are expected to be considerably smaller than the other diagrams. If $E$/$A$/$PA$ are neglected, we find two additional relations among each of the $\btos$ and $\btod$ amplitudes.

We show how the FA amplitudes can be measured through an isobar analysis of the Dalitz plots. The analysis of Ref.~\cite{Bertholet:2018tmx} cannot be applied to FA states (since $\bd \to K^0 K^0 \ok$ has no FA state), so we describe other sets of $B \to PPP$ decays that can be used to extract $\gamma$ using the FA amplitudes. Finally, we discuss how SU(3)$_F$ breaking is reduced when it is averaged over the entire Dalitz plot.

\bigskip
\noindent
{\bf Acknowledgments}: This work was financially supported by the National Science Foundation under Grant No. PHY-2013984 (BB, AH) and by NSERC of Canada (AJ, MFN, DL).

\bibliography{FAbib}
\bibliographystyle{unsrt}

\end{document}